\documentclass[fleqn,preprintnumbers,amsmath,amssymb,prb]{revtex4}
\usepackage{graphicx}
\usepackage{dcolumn}
\usepackage{bm}
\usepackage{epsf}
\newcommand{\newc}{\newcommand}
\newc{\beq}{\begin{equation}}
\newc{\eeq}{\end{equation}}
\newc{\beqa}{\begin{eqnarray*}}
\newc{\eeqar}{\end{eqnarray}}
\newc{\beqar}{\begin{eqnarray}}
\newc{\eeqa}{\end{eqnarray*}}
\newc{\bd}{\begin{displaymath}}
\newc{\ed}{\end{displaymath}}
\newc{\mbf}{\mathbf}
\begin{document}
\title{Calogero-Sutherland Model with Anti-periodic Boundary Condition: \\
Eigenvalues and Eigenstates}
\author{Arindam Chakraborty}
\author{Subhankar Ray}\email{subho@juphys.ernet.in}
\affiliation{Dept. of Physics, Jadavpur University, Calcutta 700032, India}
\author{J. Shamanna}
\email{jaya@vbphysics.net.in}
\affiliation{Physics Department, Visva Bharati University, Santiniketan 731235, India}
\date{15 April 2005}

\begin{abstract}
The $U(1)$ Calogero Sutherland Model with anti-periodic
boundary condition is studied. The Hamiltonian is reduced to
a convenient form by similarity transformation.
The matrix representation of the Hamiltonian acting on a
partially ordered state space is obtained in an upper triangular
form. Consequently the diagonal elements become the energy
eigenvalues. The eigenstates are constructed using Young diagram
and represented in terms of Jack symmetric polynomials.
The eigenstates so obtained are orthonormalized.
\end{abstract}
\maketitle

\setcounter{equation}{0}
\setcounter{page}{1}
\section{Introduction}
In the past few years several one-dimensional exactly solvable models 
in quantum many-body theory have been studied.
Of particular interest are the models with so-called long-range 
interaction having direct physical correspondences with various
condensed matter systems. The advantage of choosing 
one-dimensional systems is that many of these systems are
exactly solvable due to highly restrictive spatial
degrees of freedom. The spatial restriction in one-dimension 
introduces large quantum fluctuations resulting in failure of 
mean field approach which is known to work in higher dimensional 
systems. In addition, there exist physically interesting, realistic 
systems in one-dimension.

One of the most well known one-dimensional models
with long-range interaction is the so-called Calogero-Sutherland
model (CSM) \cite{calo62,suthJMP,suthPRA}. 
The model incorporates the idea of long-range interaction
in one-dimension assuming the interaction to fall off as
the inverse square of the distance between the particles.
The related Hamiltonian appearing in the work of Sutherland is
given by,
\beq\label{ham1}
H = - \sum_{j=1}^N \frac{\partial^2}{\partial x_j^2} + \sum_{j<k}
\frac{2 \alpha ( \alpha-1)}{d^2 (x_j-x_k)}
\eeq

The Hamiltonian is taken in units of $\hbar^2/2m$ and $d(x)$
is equal to,
\bd
d(x) = \frac{L}{\pi} \sin \left( \frac{\pi x}{L} \right)
\ed

\begin{figure}[h]
\resizebox{!}{2.0in}
{\hskip 1cm \includegraphics{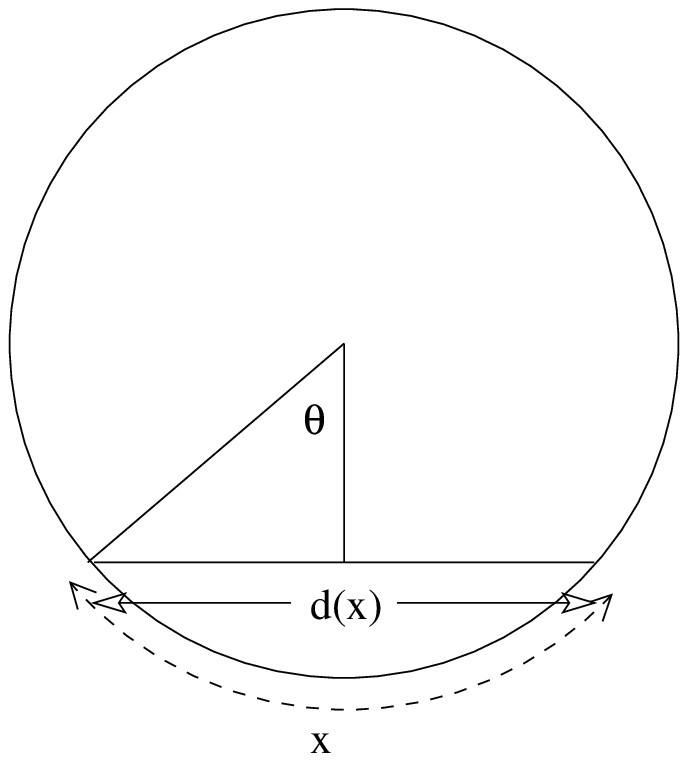}}
\resizebox{!}{2.0in}
{\hskip 1cm \includegraphics{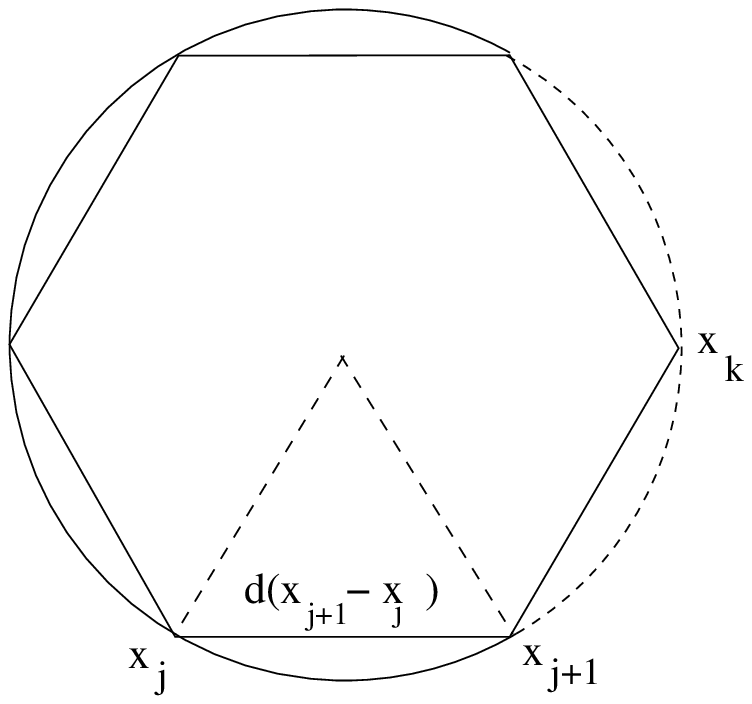}}
\caption{Interparticle distances $d(x)$ and $d(x_{j+1}-x_j)$ for particles on a circular chain}
\end{figure}
The Hamiltonian in Eq.(\ref{ham1}) describes a system
of non-relativistic quantum particles interacting with an 
inverse-square two-body potential. Here $\alpha$ is a dimensionless 
interaction parameter.
The topological representation of a one-dimensional chain is simply
a circular ring and $d(x_j-x_k)$ is the chord-distance between
the particles
at position $x_j$ and $x_k$ on a circle with circumference $L$ (length
of the chain). The spinless $U(1)$ CSM with a periodic boundary
condition
was first investigated by Sutherland\cite{suthJMP,suthPRA}.
For a spinless system the
periodic boundary condition indicates the fact that when a test particle
is transported adiabatically around the ring integral number of times, it 
does not take up any phase-factor and so the eigenfunctions retain their 
initial form.
The Hamiltonian of $U(1)$ CSM with periodic boundary
condition is usually
reduced to a simpler form by suitable gauge transformation and
the eigenvalues are obtained by operating the reduced Hamiltonian on an ordered state 
space of all
symmetric polynomials. The eigenstates are usually represented
in terms of Jack symmetric polynomials. Different generalizations of this model
have been proposed and studied \cite{habook,kojiohta,ohta}. 

In our present paper we discuss the same model with
anti-periodic boundary
condition. The general twisted boundary condition \cite{habook} appears when
a magnetic field
is allowed to pierce perpendicularly through the
plane of one-dimensional ring. As a result, when a test particle 
is transported adiabatically
around the entire system $n$ times it picks up a net phase
$\exp(i n \phi)$.
We have taken $\phi=2\pi p/q$ and $n=jq+k$, where $p$ and $q$ are
mutual primes and $j$ and $k$ are integers such that
\bd
-\infty < j < +\infty, \hskip 2cm 0 \le k \le q-1.
\ed
Consequently the interaction term becomes,
\beq
\sum _{n=- \infty}^{+\infty}\frac{\exp{(i\phi n)}}{(x+nL)^2} =
\sum _{k=0}^{q-1} \sum _{j=- \infty}^{+\infty}
\frac{1}{\left[(x+kL)+(qL)j \right] ^2}
=\sum _{k=1}^{q-1}\frac{\exp{(i 2\pi px/q)}}
{\left[(qL/\pi) \sin{\frac{\pi(x+kL)}{qL}}\right]^2}
\eeq
The effective length of the chain is thus $q L$. The interaction term
under general twisted boundary condition reduces to the so-called
anti-periodic boundary condition for $p/q = 1/2$. Hence the interaction 
term is,
\bd
\frac{\cos{(\pi x/L)}}{\frac{L^2}{\pi^2}\sin^2{(\pi x/L)}}
\ed
The final expression of the Hamiltonian takes the form
\beq\label{ham2}
H=-\sum _{j=1}^{N}\frac{\partial ^2}{\partial {x_j}^2}+2 \alpha (\alpha-1)
\frac{\pi ^2}{L^2}\sum _{j<k}\frac{\cos{[(\pi/L)(x_j-x_k)]}}{\sin^2{[(\pi/L)
(x_j-x_k)]}}
\eeq
The most important feature of such type of Hamiltonian is that
it has great physical implications in fractional exchange and 
exclusion statistics.
The so-called exchange statistics in two dimension is extremely
important
in the study of fractional quantum Hall effect (FQHE)\cite{hald81}.
In one-dimension
however, the definition of fractional exchange statistics is rather obscure
and incomplete with possible exception of the CSM. For $U(1)$
CSM the fractional
exchange statistics can be formulated in the first quantized language
by using
the one-dimensional analogue of Chern-Simon gauge field\cite{poly92}.
On the other hand,
from the same one-dimensional Hamiltonian appearing in the CSM
with periodic
boundary condition, an effective low energy model of
one-dimensional anyon
system can be constructed following the method of Luttinger
liquid theory\cite{hald82,wen90,wen91}.
For integer values of interaction parameter this one-dimensional anyon system
is equivalent to a coupled system of left and right moving edge states of
fractional quantum Hall effect. Actually there have been many suggestions
predicting a close relation of CSM with the edge states of quantum Hall
system\cite{poly89}.

Another important physical implication of CSM is in the
fractional exclusion
statistics based on the so-called generalized exclusion principle.
The formal
definition of fractional exclusion statistics is independent of spatial
dimension and is based on the structure of the Hilbert space
rather than the
configuration space. The idea of fractional exclusion statistics
was first
formulated by Haldane and applied to the elementary topological
excitations
of general condensed matter system\cite{hald91}. In the case of
$U(1)$ CSM the so-called
exclusion statistics is interpreted in terms of real , pseudo
and quasi momenta which describe the particle and hole type
excitations of the one-dimensional system\cite{ha94}. The study of
such particle-hole type excitation is extremely important to
construct the basic thermodynamic
functions of the system under consideration. In addition, the
CSM has direct
relation with many other branches of physics and mathematics
like Selberg integral (a generalization of
$\beta$-function) \cite{forr92,forr93}, $W^{\infty}$ algebra
and random matrix theory\cite{dyson62},
Jack symmetric polynomial\cite{jack} etc. which are also helpful in
turn to demonstrate many other physical properties of one-dimensional
many body system.

In our present discussion we investigate the eigenvalues and
eigenstates of Calogero-Sutherland Model with anti-periodic boundary
conditions.
The model Hamiltonian in Eq.(\ref{ham2}) is reduced to
a suitable form by
breaking the interaction term conveniently. Then, applying successive similarity
transformations, we obtain a reduced version of the same Hamiltonian.
In order to obtain the eigenvalues, we consider eigenfunctions in the
symmetric polynomial form with an ordering in the power of the variables.
The action of the Hamiltonian on a partially ordered state gives rise to a
family of mother and daughter states. We demonstrate the connection between
mother and daughter states for specific examples and their topological
representation. Finally an upper triangular representation of the Hamiltonian
matrix is obtained where the diagonal terms give the eigenvalues
of the Hamiltonian.

The eigenstates of the system is constructed in terms of the Jack
symmetric polynomials \cite{jack} which provide a complete set
of linearly independent eigenstates of the Hamiltonian. Such polynomial
eigenfunctions were initially proposed to diagonalize the so-called
Laplace-Beltrami type
differential operator and our present Hamiltonian appears as a modified
version of such operator. The Jack symmetric polynomial and all their
properties are studied from the related Young diagram representation.
The properties of the eigenstates represented by the Jack polynomials are
found to depend strongly on the interaction parameter, appearing in the
expression of the model Hamiltonian.

\section{Simplification of the model Hamiltonian}
\subsection{Gauge transformation}
Let us introduce new variable $\lambda$ defined by $\lambda=-\alpha$, 
in Eq.(\ref{ham2}). The
expression for the model Hamiltonian becomes
\beq\label{ham3}
H=-\sum _{j=1}^{N}\frac{\partial ^2}{\partial {x_j}^2}+2 \lambda
(\lambda+1)
\frac{\pi ^2}{L^2}\sum _{j<k}\frac{\cos{[(\pi/L)(x_j-x_k)]}}{\sin^2{[(\pi/L)
(x_j-x_k)]}}
\eeq
Using the transformation $\omega_j=\exp(i\pi x_j/L)$,
we get from Eq.(\ref{ham3}),
\begin{eqnarray}\label{ham41}
H=\frac{\pi^2}{L^2}\left[\sum_{j=1}^N(\omega_j\frac{\partial}{\partial
\omega_j})^2+ 4 \lambda\sum_{j<k}\left(\omega_j\frac{\partial}
{\partial\omega_j}-\omega_k\frac{\partial}{\partial\omega_k}\right)
\frac{\omega_j\omega_k}{\omega^2_j-\omega^2_k}
-4 \lambda (1-\lambda)\sum_{j<k}\frac{\omega_j\omega_k}
{(\omega_j-\omega_k)^2} \right. \\ \nonumber
\left. + 2 \lambda(1-\lambda)\sum_{j<k}\left(\frac{\omega_j^2+\omega_k^2}
{\omega_j^2 - \omega_k^2}\right)^2 
-8 \lambda(\lambda-1)\sum_{j<k}\frac{\omega_j^2+\omega_k^2}
{(\omega_j^2-\omega_k^2)^2}\;\; \omega_j\omega_k \right]
\end{eqnarray}
Let us take a similarity transformation with the
following ansatz
\begin{eqnarray}\label{simil1}
\psi=\prod_{j<k}\left(\frac{\omega_j}{\omega_k}-\frac{\omega_k}{\omega_j}\right)^\lambda \phi
\end{eqnarray}

Therefore, the Hamiltonian in units of $\pi^2/L^2$
becomes
\begin{equation}\label{ham42}
\hat{H}=\sum_{j=1}^{N}\left[\omega_j\frac{\partial}{\partial\omega_j}\right]^2
+2\lambda\sum_{j<k}\left(\frac{\omega_j+\omega_k}{\omega_j-\omega_k}\right)
\left[ \omega_j\frac{\partial}{\partial \omega_j}-\omega_k \frac{\partial}
{\partial \omega_k} \right] 
+ 4\lambda(\lambda-1)\sum_{j<k}\frac{\omega_j\omega_k}{(\omega_j-\omega_k)^2}
\end{equation}

Using again a second similarity transformation
\beq\label{simil2}
\phi=\prod_{j<k}\left(\frac{\omega_j}{\omega_k}+\frac{\omega_k}
{\omega_j}-2\right)^{\beta/2}\phi_0
\eeq
where $\beta=\beta(\lambda)$.

As a result the expression in Eq.(\ref{ham42}) becomes
\begin{eqnarray}\label{ham5}
\widetilde{H}=\sum_{j=1}^{N}\left[\omega_j\frac{\partial}{\partial\omega_j}\right]^2
+ (2\lambda+\beta(\lambda))\sum_{j<k}\left(\frac{\omega_j+\omega_k}{\omega_j-\omega_k}\right)
\left[ \omega_j\frac{\partial}{\partial \omega_j}-\omega_k \frac{\partial}
{\partial \omega_k} \right]
\end{eqnarray}

Where
\begin{eqnarray}\label{int1}
\beta(\lambda)=\frac{1}{2}\left(1-4\lambda\pm \sqrt{1+8\lambda^2}\right)
\end{eqnarray}

We get the following expression of the model Hamiltonian
\beqar\label{ham6}
\widetilde{H}=\sum_{j=1}^{N}\left[\omega_j\frac{\partial}{\partial\omega_j}\right]^2
+A(\lambda)\sum_{j<k}\left(\frac{\omega_j+\omega_k}{\omega_j-\omega_k}\right)
\left[ \omega_j\frac{\partial}{\partial \omega_j}-\omega_k \frac{\partial}
{\partial \omega_k} \right]
\eeqar

Here
\bd
A(\lambda)= \frac{1}{2}\left(1 \pm \sqrt{1+8\lambda^2}\right)
\ed
The final expression of the model Hamiltonian in
Eq.(\ref{ham6}) constitutes a commutating family of invariant
differential operators for different values of the
interaction parameter $A$. For $A=1/2,2,1$ the operator $\widetilde H$
represents the radial part of the so-called Laplace-Beltrami
type operator on Riemannian symmetric spaces. It is
observed that the operator $\widetilde H$ has two additive parts,
one representing free particle Hamiltonian and another
part containing the reduced interaction term. So the
Hamiltonian can be written in the following form
\beq\label{ham7}
\widetilde{H}=H_0+A(\lambda)H_1
\eeq

\subsection{Action of the Hamiltonian on ordered state
space}
The action of the Hamiltonian in Eq.(\ref{ham7})
can be given by the
following Bosonic basis states given by the following form

\begin{eqnarray}\label{boson}
\vert n_1....n_N\rangle=\sum_P\prod_{j=1}^{N}\omega_j^{n_{_{P_j}}}
\end{eqnarray}
The set $\{n_i \vert i =1...N\}$ can be considered as a set of Bosonic
quantum number with no restriction on their values.
The sum extends over all permutations of the integer set
$\{n_i \vert i =1...N\}$. With no loss of generality, we can introduce
an ordering $(n_1 \ge n_2 \ge \dots \ge n_N)$.
The action of $H_0$ and $H_1$ on such an ordered state space gives the following
\begin{eqnarray}\label{acham}
H_0\vert n_1....n_N\rangle=\left(\sum_{j=1}^{N}n_j^2\right)\vert n_1....n_N\rangle
\end{eqnarray}
and
\beq\label{acham1}
H_1\vert n_1....n_N\rangle=\sum_{j<k}(n_j-n_k)\left(\vert
n_1....n_N\rangle
+2 \sum_{p=1}^{n_j-n_k-1}\vert\dots,n_j-p,\dots,n_k+p,\dots\rangle \right)
\eeq

\section{The mother and daughter states}
We define $\vert n_1...n_N\rangle$ the mother state and the states
generated by squeezing a pair of quantum number by one
unit like $\vert\dots,n_j-1,\dots,n_k+1,\dots\rangle$
as the daughter states. Squeezing of mother states into daughter states is
permitted when
such squeezing retains the ordering in $n_j$, i.e., when
($n_j-n_k\ge 2$).

Levels: The family of states can be organized into levels such that the
members of a given level are
mutually not related or unreachable and the daughter of a member of a
given level always belong to a lower
level in the family. Let  $\vert u\rangle_1$  implies the highest level
mother state $\vert u'\rangle_{\mu}$ implies the family of
daughter states with
$1\le u'\le u$ and $\mu$ is an index for the state in the $u$th level.

\subsection{Generation of mother and daughter states}
Let us take $\vert6,4,3,1\rangle$ as the mother state. It is obvious that
squeezing a relevant pair of quantum numbers we
obtain a sequence of mother and daughter states. Finally we reach an
irreducible daughter state. The irreducible
state is called the ground state and is commonly denoted by
$\vert1\rangle_1$.
Table.1 shows the possible mother and daughter states
with $\vert 6, 4, 3, 1\rangle$ as the highest level mother state.
\begin{table}[h]
\caption{Mother and daughter state association}
\vskip .3cm
\begin{tabular}{c c}
\begin{tabular}{|c|c|}  \hline \hline
Mother State & Daughter State \\ \hline \hline
$\vert 6,4,3,1 \rangle = \vert6 \rangle_1$ &
$\vert5,5,3,1\rangle = \vert5 \rangle_1$ \\
& $\vert6,4,2,2 \rangle = \vert5 \rangle_2$ \\
& $\vert5,4,4,1 \rangle = \vert4 \rangle_1$ \\
& $\vert6,3,3,2 \rangle = \vert4 \rangle_2$ \\
& $\vert5,4,3,2 \rangle = \vert3 \rangle_1$ \\ \hline
$\vert5,5,3,1 \rangle = \vert5\rangle_1$ &
$\vert5,4,4,1\rangle = \vert4\rangle_1$ \\
&$\vert5,5,2,2\rangle = \vert4\rangle_3$ \\
&$\vert5,4,3,2\rangle = \vert3\rangle_1$ \\ \hline
$\vert6,4,2,2\rangle = \vert5\rangle_2$ &
$\vert6,3,3,2\rangle = \vert4\rangle_2$ \\
&$\vert5,5,2,2\rangle = \vert4\rangle_3$ \\
&$\vert5,4,3,2\rangle = \vert3\rangle_1$ \\ \hline
$\vert5,4,4,1\rangle = \vert4\rangle_1$ &
$\vert5,4,3,2\rangle = \vert3\rangle_1$ \\
&$\vert4,4,4,2\rangle = \vert2\rangle_1$ \\ \hline
\end{tabular}
&
\begin{tabular}{|c|c|}  \hline \hline
Mother State & Daughter State \\ \hline \hline
$\vert6,3,3,2\rangle = \vert4\rangle_2$ &
$\vert5,4,3,2\rangle = \vert3\rangle_1$ \\
&$\vert5,3,3,3\rangle = \vert2\rangle_2$ \\ \hline
$\vert5,5,2,2\rangle = \vert4\rangle_3$ &
$\vert5,4,3,2\rangle = \vert3\rangle_1$ \\ \hline
$\vert5,4,3,2\rangle = \vert3\rangle_1$ &
$\vert4,4,4,2\rangle = \vert2\rangle_1$ \\
&$\vert5,3,3,3\rangle = \vert2\rangle_2$ \\
&$\vert4,4,3,3\rangle = \vert1\rangle_1$ \\
& (irreducible daughter \\
& state) \\ \hline
$\vert4,4,4,2\rangle = \vert2\rangle_1$ &
$\vert4,4,3,3\rangle = \vert1\rangle_1$ \\
& (irreducible daughter \\
& state) \\ \hline
$\vert5,3,3,3\rangle = \vert2\rangle_2$ &
$\vert4,4,3,3\rangle = \vert1\rangle_1$ \\
& (irreducible daughter \\
& state) \\ \hline
\end{tabular}
\end{tabular}
\end{table}

\begin{figure}[h]
\resizebox{!}{3.0in}
{\hskip 5cm \includegraphics{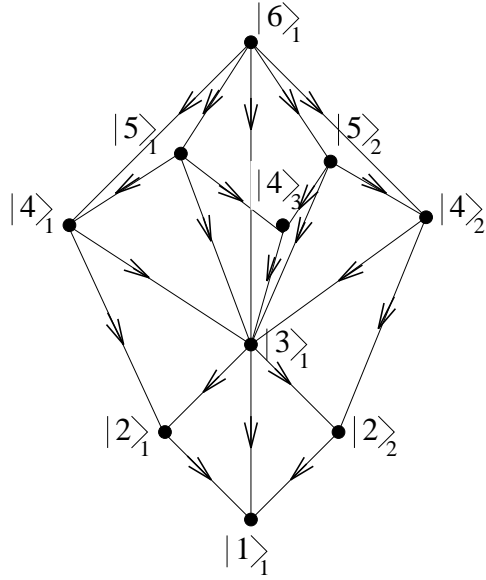}}
\caption{The topological representation of the excitation spectrum}
\end{figure}
Moreover, as the states in same level are not connected, the Hamiltonian
is diagonal in that subspace. The following
topological representation shows the connection of mother and daughter
states in the excitation spectrum. Let us define
$\nu(\eta)$ as the multiplicity of a number in a given state ket taken as
a mother state. The weight of an arrow $W$  is
given by the following equation:
\bd
W = \nu(\eta_i) \nu(\eta_j)
\ed
The following table gives the respective weights of the possible transition
from the mother to daughter states.
%
%
\begin{table}[h]
\caption{Transitions and weights for Calogero-Sutherland model
with anti-periodic boundary condition}
\vskip .3cm
\begin{tabular}{c c}
\begin{tabular}{|c|c|}  \hline \hline
Transitions & Weights (W) \\ \hline \hline
$\vert 6\rangle_1 \rightarrow \vert5 \rangle_1$ & $1 \cdot 1 = 1$ \\
$\vert 6\rangle_1 \rightarrow \vert5 \rangle_2$ & $1 \cdot 1 = 1$ \\
$\vert 6\rangle_1 \rightarrow \vert4 \rangle_1$ & $1 \cdot 1 = 1$ \\
$\vert 6\rangle_1 \rightarrow \vert4 \rangle_2$ & $1 \cdot 1 = 1$ \\
$\vert 6\rangle_1 \rightarrow \vert3 \rangle_1$ & $1 \cdot 1 = 1$ \\ \hline
$\vert 5\rangle_1 \rightarrow \vert4 \rangle_1$ & $2 \cdot 1 = 2$ \\
$\vert 5\rangle_1 \rightarrow \vert4 \rangle_3$ & $1 \cdot 1 = 1$ \\
$\vert 5\rangle_1 \rightarrow \vert3 \rangle_1$ & $2 \cdot 1 = 2$ \\ \hline
$\vert 5\rangle_2 \rightarrow \vert4 \rangle_2$ & $1 \cdot 2 = 2$ \\
$\vert 5\rangle_2 \rightarrow \vert4 \rangle_3$ & $1 \cdot 1 = 1$ \\
$\vert 5\rangle_2 \rightarrow \vert3 \rangle_1$ & $1 \cdot 2 = 2$ \\ \hline
\end{tabular}
&
\begin{tabular}{|c|c|}  \hline \hline
Transitions & Weights (W) \\ \hline \hline
$\vert 4\rangle_1 \rightarrow \vert3 \rangle_1$ & $2 \cdot 1 = 2$ \\
$\vert 4\rangle_1 \rightarrow \vert2 \rangle_1$ & $1 \cdot 1 = 1$ \\ \hline
$\vert 4\rangle_2 \rightarrow \vert3 \rangle_1$ & $1 \cdot 2 = 2$ \\
$\vert 4\rangle_2 \rightarrow \vert2 \rangle_2$ & $1 \cdot 1 = 1$ \\ \hline
$\vert 4\rangle_3 \rightarrow \vert3 \rangle_1$ & $2 \cdot 2 = 4$ \\ \hline
$\vert 3\rangle_1 \rightarrow \vert2 \rangle_1$ & $1 \cdot 1 = 1$ \\
$\vert 3\rangle_1 \rightarrow \vert2 \rangle_2$ & $1 \cdot 1 = 1$ \\
$\vert 3\rangle_1 \rightarrow \vert1 \rangle_1$ & $1 \cdot 1 = 1$ \\ \hline
$\vert 2\rangle_1 \rightarrow \vert1 \rangle_1$ & $3 \cdot 1 = 3$ \\ \hline
$\vert 2\rangle_2 \rightarrow \vert1 \rangle_1$ & $1 \cdot 3 = 3$ \\ \hline
\end{tabular}
\end{tabular}
\end{table}
\subsection{Sub-family of states}
Let us introduce the so-called subfamily of states which consist of the
highest level mother state and all her reachable
daughters. The total number of subfamily is called the dimension of the
family. Since the action of Hamiltonian on a given
state space generates states belonging to lower levels, the matrix
representation of the Hamiltonian in a partially ordered
state space is always triangular. So, the eigenvalues can be read from the
diagonal elements of the matrix and there are simple algorithms
to find the eigenvectors which we shall discuss later.
             
Finally, we see that the irreducible daughter state can be of two types
\begin{enumerate}
\item $\vert ....m, m, m,.....\rangle$
\item $\vert ....m, m, m, m-1, m-1...\rangle$
\end{enumerate}
Type.1 state corresponds to the ground state with a global Galilean
boost and Type.2 states corresponds to a single hole excitation.

\section{The energy eigenvalue of the Hamiltonian}
The energy of an eigenstate spanned by a family with the highest level
mother state$\vert n_1...n_N\rangle$ is given by
\beq\label{eigen1}
E^0(n_1,\dots,n_N)=\sum_{j=1}^N n_j^2 + A(\lambda)\sum_{j<k}(n_j-n_k)
\eeq

The off diagonal elements are found to be,
\beq\label{eigen2}
{_{\mu'}} \langle u'\vert \tilde{H} \vert u\rangle_{_\mu}= \left\{
\begin{array}{l}
\sum_{P} \left(\prod_{i\in P}W_i\right) E^1(n_1,\dots,n_N) \hfill
\hskip 1cm \forall \mu >\mu' \\
\hfill \\
0 \hfill \forall \mu < \mu'
\end{array}
\right.
\eeq
with
\bd
E^1(n_1,\dots,n_N)=2A(\lambda)\sum_{j<k}(n_j-n_k)
\ed
The sum in equation Eq.(\ref{eigen2}) is over all possible paths $P$ from
$\vert u\rangle_{\mu}$ and the product is over all weights $W_i$ of the
intermediate arrows belonging to $P$. 
The general matrix elements of the Hamiltonian is given by,
\beq\label{matrix}
{_{\mu'}} \langle u'\vert \tilde{H} \vert u\rangle_{_\mu} =
\varepsilon_{u',\mu'}^{u,\mu}
\eeq
The energy eigenvalue can be represented in terms of a free particle
momentum variable that includes the interaction term $A(\lambda)$ in a very
intricate way. The full eigenspectrum of CSM can be given by,
\beq\label{spec}
E=\frac{\hbar^2}{2m}\sum_{j=1}^N k_j^2
\eeq
Where $k_j$ is the so-called pseudo-momenta and is written as
\beq\label{psumom}
k_j=\frac{1}{L'}\left[2 \pi I_J+ \pi (A-1)\sum_{m=1}^N {\rm sgn}(k_j-k_m)\right]
\eeq
Here $L'=2L$, $L$ being the length of the chain as mentioned above.
The quantum numbers $I_j$ are now distinct half-odd integers and
are related to $n_j$s by the following equation
\bd
I_j = n_j + \frac{N+1-2j}{2}
\ed

\section{The eigenfunction of the Hamiltonian}
The polynomial form of the eigenfunctions of Eq.(\ref{ham7})
is represented
by Jack polynomials\cite{jack} which is symmetric over the variables $\omega_j$.
It can be shown in fact, that the set of Jack polynomials provide
a complete
set of linearly independent eigenfunctions of the Hamiltonian
of our present problem\cite{stanley}. The formal structure of the
eigenfunctions
can be accomplished through the following procedure.

In order to label the symmetric polynomials we choose a sequence of
nonnegative integers $\{k_i\vert i=1,\dots,N\}$, in non-increasing order
i.e $k_i \ge 0$ for all $i$
and $k_i\geq k_j$ for all $i<j$.

The sequence is known as partitions. All nonzero $k_j$ are called parts.
The total number of nonzero parts are called length which is
denoted by $\ell(\mbf{k})$. The weight of a partition is given by the
formula

\beq\label{weight}
\vert \mbf{k} \vert =\sum_{j=1}^{\ell(\mbf{k})} k_{j}
\eeq

With all the above elements we construct the so-called Young diagram
$\Delta(\mbf{k})$ giving a graphical representation of the partitions.
The Young
diagram  $\Delta(\mbf{k})$ is given by the following notation
\beq\label{eq25}
\Delta(\mbf{k}) = \{(i,j): 1 \le i \le \ell(\mbf{k}), 1 \le j \le k_i\}
\eeq

The cell labeled by $(i,j)$ is situated in the $i$th row and
the $j$th
column of the Young diagram. The diagram of $\mbf{k}$  is
therefore consist
of $\ell(\mbf{k})$ rows of length $k_j$. The conjugate of a partition
is obtained by changing all the rows of $\Delta(\mbf{k})$ to columns
in non-increasing
order from left to right i.e conjugate of $\mbf{k}$ is denoted by
$\mbf{k}'=\{k_1',k_2',\dots\}$. As an obvious consequence a partition
and it's conjugate
is related by the following relation
\beq\label{conjugate}
\sum_{i=1}^{\ell(\mbf{k})}(i-1)k_i=\sum_{i=1}^{\ell(\mbf{k}')}
\left({}^{k'_i}C_2\right)
\eeq

For a given cell $(i,j)$ of the Young diagram we define  arm length
($a(s)$)  and arm co-length ($a'(s)$) given by the following formula
\beq\label{arm}
a(s) = k_i - j, \hskip 1cm
a'(s) = j - 1
\eeq

Similarly,the so-called leg length ($l(s)$) and leg co-length  ($l'(s)$) are
given by the relations
\beq\label{leg}
l(s) = k'_j - i, \hskip 1cm
l'(s) = i - 1
\eeq

And finally the upper and lower hook lengths are given by
\beq\label{hook}
h^{\uparrow}_{\mbf{k}}(s)=l(s)+\frac{1+a(s)}{A(\lambda)}, \hskip 1cm
h^{\mbf{k}}_{\downarrow}(s)=l'(s)+1+\frac{a(s)}{A(\lambda)}
\eeq
The Young diagram of the example studied in section 3 is given in
Fig.3 with it's conjugate in Fig.4.

\begin{figure}
\resizebox{!}{2.0in}
{\includegraphics{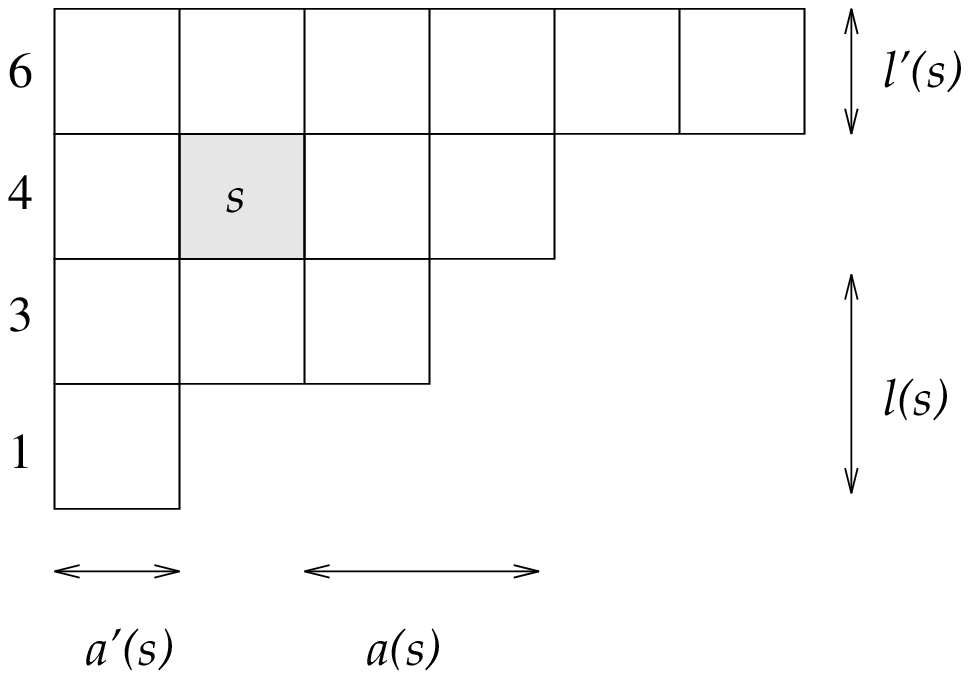} \hskip 1.5cm \includegraphics{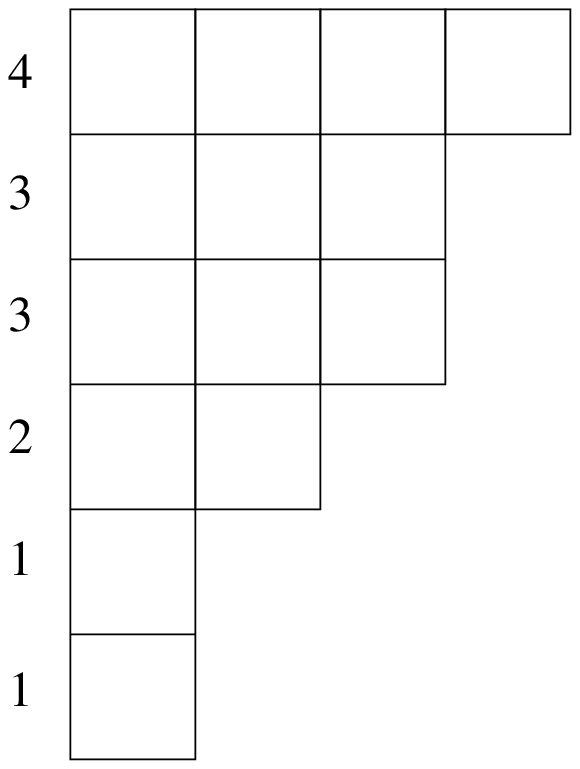}}
\caption{Young diagram for $\vert6,4,3,1\rangle$
\hskip 3cm FIG. 4: Conjugate Young diagram}

\end{figure}


With the help of above definitions we obtained the orthogonality and
normalization of the eigenstates of the Hamiltonian
in Eq.(\ref{ham7}). The Bosonic basis states $\vert \mbf{n}\rangle$
chosen earlier in Eq.(\ref{boson}) are simply the monomial symmetric 
function and the quantum number $\mbf{n}$ correspond to the partitions 
defined in the previous section. 
Though the quantum numbers may be negative
integers, the restriction on them does not affect the construction
of the
eigenstates since the CSM Hamiltonian is invariant under
a global Galilean transformation. Now the eigenstates can be given by
so-called Jack symmetric polynomials
$J^{1/A}(\omega_i \vert i=1,\dots,N)$.
If ($A=1$)
the Jack polynomial reduces to Schur functions describing the excitation
of the free Fermionic system. At ($A=0$) it becomes the
monomial symmetric
function which is nothing but the free Bosonic wave function.
For ($A=2,1/2$)
we
get the so-called zonal spherical functions. For
($A \rightarrow \infty$),
$J^{1/A}$ reduces to elementary symmetric functions.
Let us define a bilinear scalar product on the vector space of all
 symmetric function of finite degree, i.e.,
\beq\label{bilin}
\langle \eta _{\mbf{k}}\;,\;\eta_{\mbf{n}}\rangle _{1/A}=
\delta_{\mbf{k},\mbf{n}} \omega_{\mbf{k}} A^{-\ell(\mbf{k})}
\eeq
where
\bd
\eta_{\mbf{k}}=\eta_{k_1}\eta_{k_2}\eta_{k_3}\dots,
\hskip .3cm \eta_{k_\nu}=\sum_j \omega_j^{k_\nu},
\hskip .3cm \omega_{\mbf{k}}=\prod_{i\geq 1}i^{m_i} m_i!,
\hskip .3cm m_i = m_i(\mbf{k})
\ed
$\eta_{k_\nu}$ is a power sum symmetric function and
$m_i(\mbf{k})$ is the number of parts of ($\mbf{k} = i$).
Jack symmetric polynomials have the following properties\cite{macd2}
\begin{enumerate}
\item $\langle J_{\mbf{k}},J_{\mbf{n}}\rangle = \delta_{\mbf{k},\mbf{n}}
j^{A}_{\mbf{k}}$, where $j^{A}_{\mbf{k}}$ is a normalization constant.
\item $J_{\mbf{k}}=\sum_{\mbf{n}}\gamma_{\mbf{k},\mbf{n}}\phi(\mbf{n})$,
where $\gamma_{\mbf{k},\mbf{n}}=0$ unless $\mbf{k} \le \mbf{n}$.
\item If $\vert \mbf{k} \vert=p$, then $\gamma_{\mbf{k},\mbf{n}}= p!$ where
$\mbf{k} = \{\;\underbrace{1,\dots,1}_{p}\; \}$
\end{enumerate}
By the Gram-Schmidt method of orthogonalization procedure relative to the
scalar product on a ring of polynomial $J_{\mbf{k}}$ can be constructed
where $j^{A}_{\mbf{k}}$ can be given by,
\bd
j^{A}_{\mbf{k}} = \prod_{s\in\mbf{k}} h_{\mbf{k}}^{\uparrow}(s) 
h_{\downarrow}^{\mbf{k}}(s)
\ed

Jack polynomials are also orthogonal in the following manner,
\beqar\label{ortho}
\langle\mbf{k}\vert\mbf{n}\rangle_{1/A}
=&& C_N^2 \prod_{j=1}^N \left( \int _0^{L'} dx_j
\overline{J_{\mbf{k}}^{1/A}}
(\omega _1,\omega _2,\dots,\omega _N) \right. \nonumber \\
&& \left. J_{\mbf{n}}^{1/A}(\omega _1,\omega _2,\dots,\omega _N)
\cdot \prod_{i<j} \vert \omega_i-\omega_j \vert^{2 A} \right) \nonumber \\
=&& C_N^2 j^{A}_{\mbf{k}} \prod_{s\in\mbf{k}} \frac{N+a'(s)/A -l'(s)}
{N+ (a'(s)+1)/A - (l'(s)+1)} \delta_{\mbf{k},\mbf{n}}
\eeqar
where,
\bd \omega_j = e^{\frac{2 \pi x_j}{L'}}, \hskip 1cm
C_N^2 =  \left(\frac{1}{L'}\right)^N \frac{\Gamma^N(1+A)}{\Gamma(1+A N)}, \hskip
 1cm
L' = 2 L
\ed
and bar over $\overline{J_{\mbf{k}}^{1/A}}$ implies
complex conjugation.

\section{Conclusion}
In this article we obtain the energy eigenvalues and the eigenstates of
the CSM with anti-periodic boundary condition.
The model Hamiltonian is reduced to a convenient form by means of
similarity transformations. Finally, we obtain an upper triangular
representation of the Hamiltonian in terms of a partially ordered basis.
The eigenstates are chosen to be symmetric polynomials
known as Jack symmetric polynomials. These eigenfunctions are constructed 
by studying the Young diagram representation.
The orthogonality and normalization of the eigenfunctions are also discussed.

\section*{Acknowledgment}
AC wishes to acknowledge the Council of Scientific and
Industrial Research, India (CSIR) for fellowship support.


\end{document}